\documentclass{article}
\usepackage[tbtags]{amsmath}
\usepackage{amsfonts,verbatim,amssymb}
\usepackage{mathrsfs}
\usepackage{amsthm}
\usepackage{showlabels}
\allowdisplaybreaks

\theoremstyle{plain}
\newtheorem*{theorem}{Theorem}

\theoremstyle{definition}

\begin{document}


\date{}

\title{A Logical Proof of the Free-Will Theorem}

\author{Iegor Reznikoff\\[1ex]
Professor Emeritus, Departement de Philosophie,\\
Universit\'e de Paris-Ouest, 92001 Nanterre, France\\
\textit{E-mail}: dominiqueleconte@yahoo.fr}



\maketitle



 This proof refers to the known paper by J.~Conway and S.~Kochen which appeared
in the ``Notices of the AMS,'' in February 2009 (Vol.~56/2, p.~226).
 Without loss of generality, we reduce the problem to its relevant main logical
structure.
 We assume that with a particle is associated a finite number~$n$
of ``directions'' or ``components'' ($n\ge2$); at time~$t$, on each component~$i$,
the particle either verifies the property or does not: we write
$P(t,i)=0$ or $P(t,i)=\nobreak1$, respectively. 
 At any time~$t$, a component~$i$ of the particle ``can be observed and measured'';
we write $A(t,i)$ if the component~$i$ is observed at time~$t$
(formally, $A(t,i)$ can be considered as a predicate symbol). For the sake of coherence
with the informal observational meaning, we may introduce the axiom that,
at a given time, only one component can be observed:
\medskip

1)
\qquad
$\forall t,\ \forall i,j,\quad [A(t,i)\wedge A(t,j)\to (i=j)]$.
\medskip

Formally, however, this axiom is not necessary for our purpose.
The relevant axioms are:
\medskip

2) At any time~$t$, if the component~$i$ is observed, its value is~$0$:
\medskip

\noindent
$\hphantom{A}$\hfill$\forall t,\ \forall i,\quad [A(t,i)\to P(t,i)=0]$.\hfill $\hphantom{A}$
\medskip

3) At a given time~$t$, not all components have the value~$0$:
\medskip

\noindent
$\hphantom{A}$\hfill$\forall t,\ \urcorner\,\forall i,\quad P(t,i)=0$.\hfill$\hphantom{A}$
\medskip

 Axiom~2) corresponds to the SPIN axiom of the Conway--Kochen paper and refers
to the measurement of spin operators of a quantum particle, which, in three
orthogonal directions, gives the values $1,0,1$ in some order;
in this case, we write $P(t,i)=0$, where $i$
is one of the $n=33$ Peres directions of orthogonal triples.
 Axiom~2) then means that the measurement in the direction~$i$ at time~$t$
gives the expected answer.
 Axiom~3) is a structural axiom corresponding to the Bell
or Kochen--Specker theorem (or paradox): it is impossible
to have a positive answer in all the 33 directions.
 Of course, if all components could be observed at the same time,
axioms~2) and~3) would be contradictory.
 This is a reason for the first axiom.

 Now---in order to speak of the freedom of a particle---we assume that
there is a theory~$F$ expressed in the language of physics, mathematics, and logic
in which the notions indicated above are defined and axioms~2) and~3) for the particle
are included and which gives a deductive frame for the notion of freedom;
namely, we say that a \textit{particle is free} if, at any time~$t$,
at least the value on one of its components cannot be predetermined
(its value proved) by events at time $t'<t$ in the theory~$F$:
\medskip

(a)\qquad $\forall t,\ \exists i,\quad (F(t')\nvdash P(t,i)=0\quad
\text{and}\quad F(t')\nvdash P(t,i)=1)$,
\medskip

\noindent
where in~$F$, for that matter, time variables~$t'$, different from~$t$,
are such that $t'<t$ (we write briefly $t'$ for all time variables
different from~$t$, and~$t$ is the time at which the value on the component
is to be determined); moreover, only events that happen at time $t'<t$
appear in~$F$.
 This is easy to understand, but complicated to define.
 However, logically, formally, we need not do this here.

 Finally, we say that an observation (or measurement) or, better,
an \textit{observer}~$A$ \textit{is free} if, at each time~$t$,
$A$ is free to observe any component~$i$.
 This apparently only an intuitive notion can be treated in purely
logical terms: in $A(t,i)$, the variable~$i$ is free or independent
of the variable~$t$ (they are variables of different kind) and,
moreover, $A(t,i)$ cannot be deduced from events at time $t'<t$;
thus, with conditions as in~(a) above:
\medskip

(b)\qquad $\forall t,\ \forall i,\quad (F(t')\nvdash A(t,i))=0\quad
\text{and}\quad F(t')\nvdash\,\urcorner\, A(t,i))$.

\begin{theorem}
If the theory is consistent and the observer~$A$ is free, then
the particle is also free.
\end{theorem}

\begin{proof}
If the particle is not free at time~$t$, then, from~$(a)$,
we get that, for each component, its value at this time
can be determined, i.e.,
\medskip

$\hphantom{A}$\hfill$\forall i,\quad (F(t')\vdash P(t,i)=0\quad
\text{or}\quad F(t')\vdash P(t,i)=1)$.\hfill$\hphantom{A}$
\medskip

\noindent
Since, by axiom~3), not for all~$i$, P(t,i)=0, there exists an~$i$
such that
\medskip

$(*)$\qquad $F(t')\nvdash P(t,i)=0$.
\medskip

\noindent
But, however, it cannot be shown, positively, for this component~$i$ that
\medskip

(c)\qquad $F(t')\vdash P(t,i)=1$.
\medskip

\noindent
Indeed:
\medskip

\textsl{Informal proof.} Since only events that happen at time~$t'<t$
appear in $F(t')$ and the observer~$A$ is free, it follows that, at time~$t$,
this observer is free to observe~$i$; then $A(t,i)$ is true and, from axiom~2),
we obtain $P(t,i)=0$, which contradicts~(c) above. Therefore
\medskip

$(**)$\qquad $F(t')\nvdash P(t,i)=1$.
\medskip

\textsl{Formal proof.} From~(c) we get
\medskip

(d)\qquad $F(t'),A(t,i)\vdash P(t,i)=1$,
\medskip

\noindent
but $A(t,i)$, together with axiom~2), gives $P(t,i)=0$,
and hence also
\medskip

(e)\qquad $F(t'),A(t,i)\vdash P(t,i)=0$,
\medskip

\noindent
This means that $F(t')$ and $A(t,i)$ are contradictory, and hence
$F(t')\vdash\, \urcorner\, A(t,i)$, which, in turn, contradicts the second half
of definition~(b) above (because the observer is supposed to be free).
Finally, we see that~(c) is contradictory, which implies~$(**)$.

But~$(*)$ and~$(**)$ together mean precisely that the particle is free at time~$t$
on the component~$i$, and this contradicts the hypothesis of the
\textit{reductio ad absurdum} proof. Thus, we have proved that,
at any time~$t$, there is a component of the particle whose value cannot
be predetermined in the theory.
\end{proof}

By an elementary combinatorial argument, the result can be strengthened
by requiring in the definition of a free particle that always at least the values
on \textit{two} components are not determined.
\bigskip

$\hphantom{A}$\hfill Helsinki, 18 June, 2010


\end{document}